\documentclass[prl,aps,twocolumn,superscriptaddress,a4paper,sort&compress,balancelastpage,10pt]{revtex4-2}

\usepackage{amsmath,amssymb,calrsfs}
\usepackage{graphicx}
\usepackage[colorlinks=true,citecolor=blue,urlcolor=blue,linkcolor = blue]{hyperref}  
\usepackage{dcolumn}
\usepackage{bm}
\usepackage{verbatim}
\usepackage{units}
\usepackage{upgreek}
\usepackage{multirow}

\usepackage[T1]{fontenc}
\usepackage[utf8]{inputenc}

\usepackage[dvipsnames]{xcolor}

\usepackage[normalem]{ulem}

\usepackage{booktabs}

\newcommand{\Sec}[1]{{\textit{#1.---}}}

\newcommand{\TITLE}{Universality in strongly interacting bosonic clusters}

\begin{document}

\title{\TITLE}

\author{L. Madeira}
    \email[Corresponding author: ]{lmadeira@ectstar.eu}
    \affiliation{INFN-TIFPA Trento Institute of Fundamental Physics and Applications, Via Sommarive 14, I-38123 Trento, Italy}
	\affiliation{European Centre for Theoretical Studies in Nuclear Physics and Related Areas (ECT*), 
    Fondazione Bruno Kessler, Strada delle Tabarelle 286, I-38123 Trento, Italy}

\author{F. Pederiva}
	\affiliation{INFN-TIFPA Trento Institute of Fundamental Physics and Applications, Via Sommarive 14, I-38123 Trento, Italy}
	\affiliation{Physics Department, University of Trento, Via Sommarive 14, I-38123 Trento, Italy}

\author{U. van Kolck}
	\affiliation{European Centre for Theoretical Studies in Nuclear Physics and Related Areas (ECT*), 
    Fondazione Bruno Kessler, Strada delle Tabarelle 286, I-38123 Trento, Italy}
	\affiliation{Université Paris-Saclay, CNRS/IN2P3, IJCLab, 91405 Orsay, France}
	\affiliation{Department of Physics, University of Arizona, Tucson, AZ 87521, USA}
	
\date{\today}

\begin{abstract}
We develop an effective field theory (EFT) for strongly interacting bosonic clusters, using $^4$He as a paradigmatic example of universality in systems with large scattering length. At leading order (LO), two- and three-body zero-range interactions are entirely determined by the dimer and trimer ground-state energies. We show that ground-state energies for up to $N=15$ particles converge to cutoff-independent limits with extrapolation coefficients of natural size. At next-to-leading order (NLO), corrections stemming from the two-body interaction range and a four-body force, calibrated to the tetramer ground-state energy, reduce cutoff sensitivity. Close agreement with results from a realistic potential is found at LO and improved at NLO, demonstrating systematic convergence with few parameters at each order. The resulting EFT is directly applicable to larger clusters and bulk helium. 
\end{abstract}

\maketitle

\Sec{Introduction} Strongly interacting systems characterized by large two-body scattering lengths exhibit universal behavior~\cite{Braaten:2004rn,Naidon:2016dpf}. Realizations include ultracold atoms near Feshbach resonances~\cite{Chin:2010crf} and nuclear few- and many-body systems~\cite{Hammer:2019poc,Kievsky:2021ghz}. Although their 
energy and length scales differ by orders of magnitude, ratios of physical quantities in few-body systems take similar values regardless of 
short-distance details. Increased understanding of any member of this class therefore yields insight across several areas of physics.

Atomic $^4$He provides a paradigmatic realization of this class of strongly interacting systems. Numerous high-precision interatomic potentials have been developed~\cite{Aziz:1979,Aziz:1991,Janzen:1995,Janzen:1997,Korona:1997,Przybytek:2010zz} and used to compute properties of helium clusters~\cite{Pandharipande:1983xez,Kievsky:2011ut,Hiyama:2011ge,Hiyama:2012cj,Hiyama:2014kia,Kievsky:2014dua,Kievsky:2020sni,Yates:2022mxk,Recchia:2022jih} and bulk helium~\cite{Kalos:1981zz,Ceperley:1995zz,Moroni:2000}. The large two-body scattering length and associated Efimov physics~\cite{Efimov:1970zz,Efimov:1971,Kunitski:2015qth} generate universal features largely independent of the underlying potential. A systematic description of universality and deviations from it is provided by
an effective field theory (EFT) with zero-range interactions~\cite{Bedaque:2002mn}. The same EFT framework employed for atomic $^4$He can be applied to other systems~\cite{Hammer:2019poc}, such as light and halo nuclei, albeit with additional challenges arising from the exclusion principle.

The EFT expands the interaction in Dirac delta functions and their derivatives~\cite{Bedaque:2002mn}, analogous to the multipole expansion in classical electrodynamics. Universality is captured explicitly at leading order (LO) with two parameters associated with the small energies of the dimer and trimer~\cite{Bedaque:1998kg,Bedaque:1998km}. Corrections to universality enter at subleading orders, starting at next-to-leading order (NLO) with two additional parameters related to the two-body interaction range and the tetramer energy~\cite{Bazak:2018qnu}. While in the two-body sector the EFT yields an $S$ matrix that reproduces, order-by-order, the effective range expansion, energy-dependent pseudopotentials, and generic boundary conditions at the origin~\cite{vanKolck:1998bw}, its Hamiltonian dynamics can be solved through the Schr\"odinger equation in the two-, few-, and many-body sectors.

After it
is calibrated in the few-body sector, 
the EFT is used to predict properties of larger systems.
A central challenge in many areas of physics~\cite{Hammer:2019poc}, 
which we address here, is the extent to which universality determines the properties of systems with an increasing number of particles.

\begin{figure*}[tb]
    \centering
    \includegraphics[scale = 1.0]{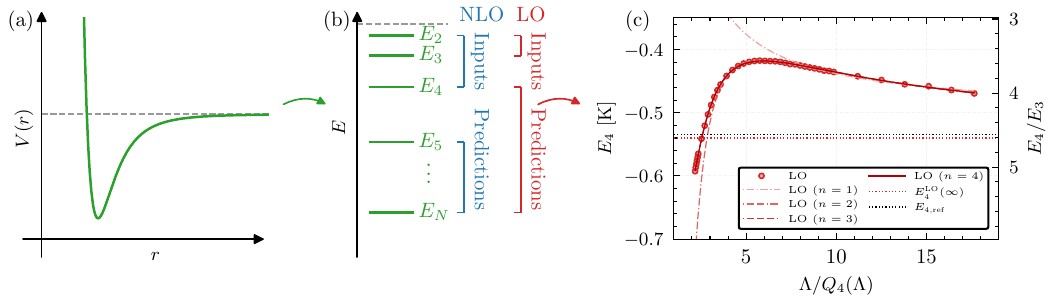}
    \caption{(a) Schematic interatomic potential used to generate reference data. (b) EFT calibration and prediction: at LO the dimer and trimer energies ($E_2$,$E_3$) are taken as input, yielding predictions for $N\geqslant 4$; at NLO the inputs are the two-body effective range and $E_2$, $E_3$, $E_4$, yielding predictions for $N\geqslant 5$. (c) LO tetramer energy $E_4$ (in K and in units of the trimer energy $E_3$) versus the dimensionless ratio $\Lambda/Q_4(\Lambda)$, where $Q_4(\Lambda)$ is computed through Eq.~(\ref{eq:Q_N}) for each $E_4(\Lambda)$. Circles show DMC results obtained with the LO interaction; statistical uncertainties are smaller than the symbols. Curves are fits to Eq.~(\ref{eq:ext_powers}) with truncation orders $n=1$--4 (see legend); the $n=3$ and $n=4$ curves are indistinguishable on the scale shown, indicating convergence. The horizontal lines denote the infinite-cutoff extrapolation $E^{\rm LO}_4(\infty)$ using $n=4$ (red) and the HFDHE2 reference value (black).}
    \label{fig:vr_levels_N4}
\end{figure*}

Determination of the EFT parameters is done by matching to experimental data or to a shorter-distance theory. 
Figure~\ref{fig:vr_levels_N4}(a,b) schematically illustrates 
the procedure in the case of helium clusters where we match to a potential model. Successful LO and NLO descriptions have been achieved but so far 
largely limited to few-body systems~\cite{Bedaque:1998kg,Bedaque:1998km,Braaten:2002sr,Braaten:2002jv,Platter:2004he,Platter:2006ev,Ji:2012nj,Bazak:2016wxm,Bazak:2018qnu,Contessi:2023yoz,Wu:2023mhg,Bazak:2025usn,Wu:2026pjt}.
In this Letter, we show that the EFT accurately predicts properties of much larger clusters containing up to $N=15$ helium atoms. 
We deploy quantum Monte Carlo methods to compute ground-state energies of helium clusters from LO and NLO interactions. The systematic order-by-order improvement 
we find shows that the gross properties of helium droplets are universal. This extension opens the door to a systematic description of bulk helium, a system with remarkable properties that remains liquid down to zero temperature at low pressures, where it becomes superfluid. 

\Sec{Theory} The ground-state energies of the $N$-particle clusters $E_N$ are obtained by solving the 
Schr\"odinger equation for $N$ nonrelativistic particles of mass $m$ with the Hamiltonian
\begin{equation}
\label{eq:H}
H=-\frac{\hbar^2}{2m}\sum_{i=1}^N \nabla_i^2 + V(\bm{R}),
\end{equation}
where $\bm{R}$ denotes the $3N$ particle coordinates. We employ the Diffusion Monte Carlo (DMC) method~\cite{Foulkes:2001zz}, which propagates a trial wave function $\psi_T$ in imaginary time $\tau$,
\begin{equation}
\label{eq:propagation}
\psi(\tau)=e^{-(H-E_T)\tau}\psi_T,
\end{equation}
with an energy offset $E_T$. In the limit $\tau\to\infty$, the propagation projects onto the ground state within controllable statistical uncertainties. The trial wave function includes two-body (and, where required, three-body) correlations, following Refs.~\cite{Carlson:2017txq,Madeira:2021qcz}.

Data on $^4$He clusters are required both to calibrate the EFT and to benchmark its predictions. Experimental measurements of few-body $^4$He systems are very challenging~\cite{Grisenti:2000zz,Voigtsberger:2014,Kunitski:2015qth,Zeller:2016mwo}, and ground-state energies are not available up to $N=15$. Therefore, we generate reference data by solving Eq.~(\ref{eq:H}) with a realistic interatomic $^4$He potential. The specific choice of potential is unimportant, since our goal is to construct an EFT from few-body properties and then predict the spectrum of a strongly interacting system. We adopt the HFDHE2 potential~\cite{Aziz:1979}, widely used in cluster calculations~\cite{Pandharipande:1983xez,Hiyama:2011ge,Kievsky:2014dua,Kievsky:2020sni,Yates:2022mxk}, and known to reproduce the $^4$He equation of state in agreement with experiment~\cite{Kalos:1981zz}. We generate energies $E_{N,{\rm ref}}$ using this realistic potential yielding results consistent with previous calculations~\cite{Pandharipande:1983xez,Yates:2022mxk}.

Observables such as $E_N$ are obtained in the EFT as an expansion in $Q_N {\mathcal{R}}/\hbar$, where $\mathcal{R}$ is a measure of the range of the underlying interaction and 
\begin{equation}
\label{eq:Q_N}
Q_N=\sqrt{-2mE_N/N}
\end{equation}
is an estimate of the typical momentum of the $N$-particle system. Because the EFT potentials are singular, they require regularization with an arbitrary function, which we take to be a Gaussian, characterized by a momentum cutoff $\Lambda$. Independence from this choice is ensured by renormalization, where non-negative powers of $\Lambda$ are eliminated from observables at a given order and negative powers, at higher orders. We verify that this is achieved by fitting our finite-cutoff results at order $\nu$ in the EFT expansion with
\begin{equation}
\label{eq:ext_powers}
E_N^{(\nu)}(\Lambda)=E_{N}^{(\nu)}(\infty)\left[1+\sum_{i=1}^{n} c_i^{(\nu)}\left(\frac{Q_N^{(\nu)}(\Lambda)}{\Lambda}\right)^i\right].
\end{equation}
The momentum 
$Q_N^{(\nu)}$ entering Eq.~(\ref{eq:ext_powers}) is computed from Eq.~(\ref{eq:Q_N}) using the energy evaluated at the same cutoff, $E_N^{(\nu)}(\Lambda)$. Since $Q_N^{(\nu)}(\Lambda)$ and its infinite-cutoff limit differ by a relative factor of ${\cal O}(Q_N/\Lambda)$, this choice only amounts to a reshuffling of the coefficients.  
A controlled EFT expansion requires natural coefficients 
$c_i^{(\nu)}={\cal O}[(Q_N{\mathcal{R}}/\hbar)^\nu]$, ensuring systematic suppression of higher-order terms.
We take $E_{N}^{(\nu)}(\infty)$ as the EFT prediction at order $\nu$; in the following, $\nu=0$ and $\nu=1$ correspond to LO and NLO (i.e., LO plus the first-order correction), respectively.

\Sec{Leading order} At LO, the interaction consists of two- and three-body potentials~\cite{Bazak:2018qnu},
\begin{eqnarray}
V^{\rm LO}(\bm{R};\Lambda)
&=&V_2^{\rm LO}(\Lambda) \sum_{i<j} e^{-r_{ij}^2 \Lambda^2/4\hbar^2}\nonumber\\
&+&V_3^{\rm LO}(\Lambda) \sum_{i<j<k} e^{-(r_{ij}^2+r_{ik}^2+r_{jk}^2) \Lambda^2/6\hbar^2}.
\label{eq:LO}
\end{eqnarray}
The ``low-energy constants'' (LECs) $V_2^{\rm LO}(\Lambda)$ and $V_3^{\rm LO}(\Lambda)$ are determined at each $\Lambda$ by fitting the $^4$He dimer and trimer energies, respectively, as detailed in the Supplemental Material \cite{sup}.

The first nontrivial EFT prediction is the LO tetramer energy, shown in Fig.~\ref{fig:vr_levels_N4}(c). The dependence on the largest cutoffs is already captured by the $n=1$ term in Eq.~(\ref{eq:ext_powers}), while including higher powers extends the description to smaller cutoffs.
The extrapolated infinite-cutoff limit using $n=4$ agrees 
with the HFDHE2 result within about 1\%.
For comparison, Refs.~\cite{Bazak:2016wxm, Bazak:2018qnu} find agreement at the 3-10\% level using a different EFT calibration to another potential.

\begin{figure}[tb]
    \centering
    \includegraphics[scale = 0.84]{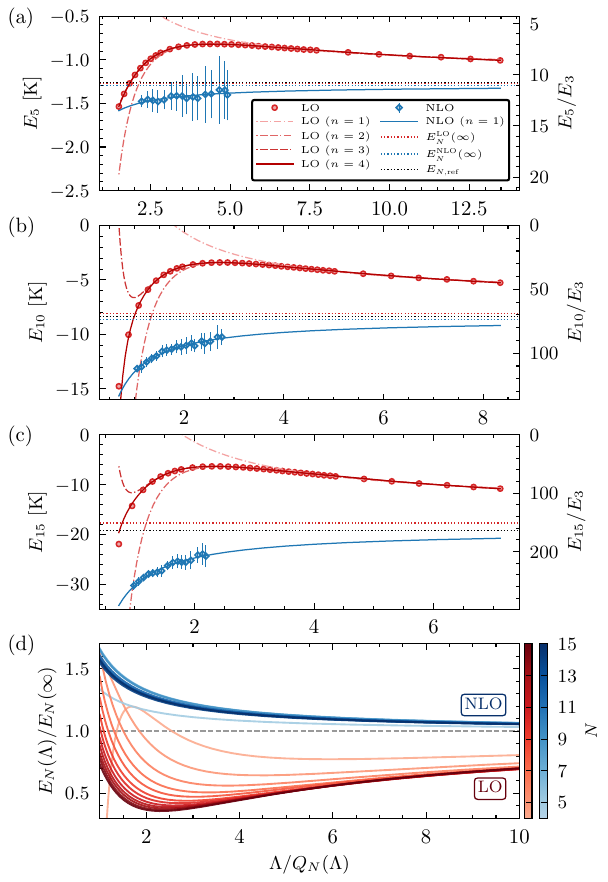}
    \caption{
    Dependence on the cutoff $\Lambda$ [through the dimensionless ratio $\Lambda/Q_N(\Lambda)$] of cluster energies $E_N$ (in K and in units of $E_3$) for (a) $N=5$, (b) $N=10$, and (c) $N=15$. LO 
    symbols are as in Fig.~\ref{fig:vr_levels_N4}(c). NLO results are shown as blue diamonds together with the $n=1$ fit of Eq.~(\ref{eq:ext_powers}) and the corresponding NLO infinite-cutoff extrapolation.
(d) Scaled energies $E_N(\Lambda)/E_N(\infty)$ 
for LO (red) and NLO (blue) obtained from Eq.~(\ref{eq:ext_powers}) with $n=4$ and 1, respectively.
} 
    \label{fig:N5_N10_N15}
\end{figure}

The same procedure yields predictions for all larger clusters $5\leqslant N \leqslant 15$. For conciseness, Figs.~\ref{fig:N5_N10_N15}(a-c) show representative results for $N=5$, 10, and 15, while analogous plots for the remaining systems are provided in the Supplemental Material~\cite{sup}. The behavior observed for the tetramer persists at larger $N$: the EFT 
fit of Eq.~(\ref{eq:ext_powers}) with $n=4$ truncation accurately describes the cutoff dependence across the entire range of cluster sizes considered. 
The relevant fit parameters 
are given in Table~\ref{tab:LO_NLO}, while a systematic analysis of the $Q_N(\Lambda)/\Lambda$ expansion is provided in the Supplemental Material~\cite{sup}. The $\{c_1^{\rm LO}\}$ are natural and of comparable magnitude, demonstrating the predictive consistency of the EFT expansion,
with indications of saturation at the largest clusters studied.

\begin{table}[bt]
\centering
\caption{Parameters of Eq.~(\ref{eq:ext_powers}) for LO ($n=4$) and NLO ($n=1$).
For LO, only the leading coefficient $c_1$ is shown.
The last two columns give the ratio of the infinite-cutoff extrapolated energy 
$E_N(\infty)$ to the corresponding HFDHE2 value, $E_{N,\mathrm{ref}}$.}
\label{tab:LO_NLO}
\begin{tabular}{c|cc|cc|cc}
\toprule
$N$ & \multicolumn{2}{c|}{$c_1$} 
 & \multicolumn{2}{c|}{$ E_N(\infty)$ \![K]} 
 & \multicolumn{2}{c}{$E_N(\infty)/E_{N,\mathrm{ref}}$} \\
& LO & NLO & LO & NLO & LO & NLO\\
\midrule
\addlinespace
4  &  $-2.90(6)$ &           & $-0.540(2)$  &            & 1.010(3) & \\
5  &  $-3.55(4)$ & $0.34(7)$ & $-1.271(4)$  & $-1.29(2)$ & 1.008(4) & 1.03(2) \\
6  &  $-3.75(3)$ & $0.58(5)$ & $-2.272(6)$  & $-2.28(4)$ & 1.006(3) & 1.01(2) \\
7  &  $-3.81(3)$ & $0.66(5)$ & $-3.49(1)$   & $-3.51(6)$ & 1.001(3) & 1.01(2) \\
8  &  $-3.76(3)$ & $0.65(5)$ & $-4.85(2)$   & $-5.0(1)$  & 0.984(3) & 1.02(2) \\
9  &  $-3.82(3)$ & $0.65(6)$ & $-6.44(3)$   & $-6.6(2)$  & 0.983(4) & 1.01(2) \\
10 &  $-3.77(3)$ & $0.56(3)$ & $-8.07(3)$   & $-8.6(1)$  & 0.967(4) & 1.03(1) \\
11 &  $-3.76(3)$ & $0.58(3)$ & $-9.86(5)$   & $-10.6(2)$ & 0.958(4) & 1.03(1) \\
12 &  $-3.72(3)$ & $0.61(4)$ & $-11.70(6)$  & $-12.5(2)$ & 0.947(5) & 1.01(2) \\
13 &  $-3.64(6)$ & $0.58(4)$ & $-13.6(1)$   & $-14.6(3)$ & 0.932(7) & 1.00(2) \\
14 &  $-3.62(6)$ & $0.55(5)$ & $-15.6(1)$   & $-17.1(4)$ & 0.925(8) & 1.01(2) \\
15 &  $-3.60(4)$ & $0.57(4)$ & $-17.7(1)$   & $-19.2(4)$ & 0.919(5) & 1.00(2) \\
\bottomrule
\end{tabular}
\end{table}

Compared with the HFDHE2 reference values, the LO predictions agree within a few percent for small clusters and remain within 10\% even for the largest systems considered, demonstrating that the EFT already provides a quantitatively accurate description of the system at LO.


\Sec{Next-to-Leading Order} At NLO, the corrections are evaluated perturbatively using LO ground-state wave functions. The NLO interaction contains two-, three-, and four-body terms~\cite{Bazak:2018qnu},
\begin{align}
&V^{\rm NLO}(\bm{R};\Lambda)=
V_{2,0}^{\rm NLO}(\Lambda) \sum_{i<j} e^{-r_{ij}^2 \Lambda^2/4\hbar^2}\nonumber\\
&+ V_{2,2}^{\rm NLO}(\Lambda) \sum_{i<j} (r_{ij} \Lambda/\hbar)^2 e^{-r_{ij}^2 \Lambda^2/4\hbar^2}\nonumber \\
&+ V_3^{\rm NLO}(\Lambda) \sum_{i<j<k} e^{-(r_{ij}^2+r_{ik}^2+r_{jk}^2) \Lambda^2/6\hbar^2}\nonumber \\
&+ V_4^{\rm NLO}(\Lambda)  \sum_{i<j<k<l}
e^{-(r_{ij}^2+r_{ik}^2+r_{il}^2+r_{jk}^2+r_{jl}^2+r_{kl}^2)
\Lambda^2/4\hbar^2},
\label{eq:NLO}
\end{align}
introducing four NLO LECs. As detailed in the Supplemental Material \cite{sup}, distorted-wave perturbation theory (DWPT) is used
to reproduce 
the two-body effective range and the dimer, trimer, and tetramer energies. DWPT accounts for renormalization and a systematic EFT expansion \cite{Hammer:2019poc}.

Imaginary-time propagation in DMC yields exact results, within statistical uncertainties, for observables that commute with the Hamiltonian. 
The NLO correction, however, involves matrix elements of operators [Eq.~(\ref{eq:NLO})] that do not commute with $H^{\mathrm{LO}}$. While such quantities can be estimated within DMC using specialized estimators, we instead employ Variational Monte Carlo (VMC)~\cite{Foulkes:2001zz}, which evaluates non-commuting operators directly but depends on the quality of the trial wave function. We adopt neural-network wave functions that accurately describe strongly interacting ground states~\cite{Carleo:2016svm,Lovato:2022tjh,Gnech:2021wfn}. The implementation builds on the framework of Refs.~\cite{Freitas:2023yed,Freitas:2024}, extended here to include regularized two- and three-body contact interactions and to evaluate the NLO matrix elements. As validation, LO energies computed with VMC are found to be in excellent agreement with DMC results.

We computed NLO predictions for $5\leqslant N \leqslant 15$ clusters, with sample results shown in Figs.~\ref{fig:N5_N10_N15}(a-c).  
They 
lie systematically closer to the HFDHE2 reference than LO even before extrapolation, demonstrating the expected EFT pattern of order-by-order improvement. The cutoff range considered at NLO is smaller than at LO for two reasons. First, for a given $\Lambda$, the ratio $\Lambda/Q_N(\Lambda)$ is reduced at NLO because the clusters are more bound than at LO [cf. Eq.~(\ref{eq:Q_N})]. Second, the variance of the NLO matrix elements increases with the cutoff as the Gaussian regulators become narrower, requiring larger Monte Carlo samples to maintain fixed precision. This reflects a technical limitation rather than a deficiency of VMC or EFT and can be mitigated with increased computational effort. Owing to the larger uncertainties and reduced cutoff interval, the extrapolation of Eq.~(\ref{eq:ext_powers}) is performed using only the first term ($n=1$). The NLO coefficients, shown in Table~\ref{tab:LO_NLO}, are of comparable magnitude and
5-10 times smaller than the LO ones, indicating faster convergence with increasing cutoff.
As was the case for LO, the coefficients $\{c_1^{\rm NLO}\}$ also show signs of saturation as the cluster size increases.

Despite 
the technical limitation, the NLO predictions lie within no more than 5\% 
of the HFDHE2 reference, even for $N=15$. The NLO improvement, which had 
been 
shown only for $N=5,6$ for a different calibration with another reference~\cite{Bazak:2018qnu}, is now seen to hold beyond few-body systems. 

\Sec{Order-by-order convergence} The impact on EFT convergence is illustrated in Fig.~\ref{fig:N5_N10_N15}(d), which shows the scaled energies $E_N(\Lambda)/E_N(\infty)$ from Eq.~(\ref{eq:ext_powers}). At LO, the cutoff behavior is visibly $N$-dependent, most pronounced for $N=4$ and weaker for larger clusters. 
The smaller $c_1$ coefficients at NLO mean  
the curves  
have substantially reduced spread, indicating improved convergence and reduced cutoff sensitivity.

\begin{figure}[tb]
    \centering
    \includegraphics[scale = 0.99]{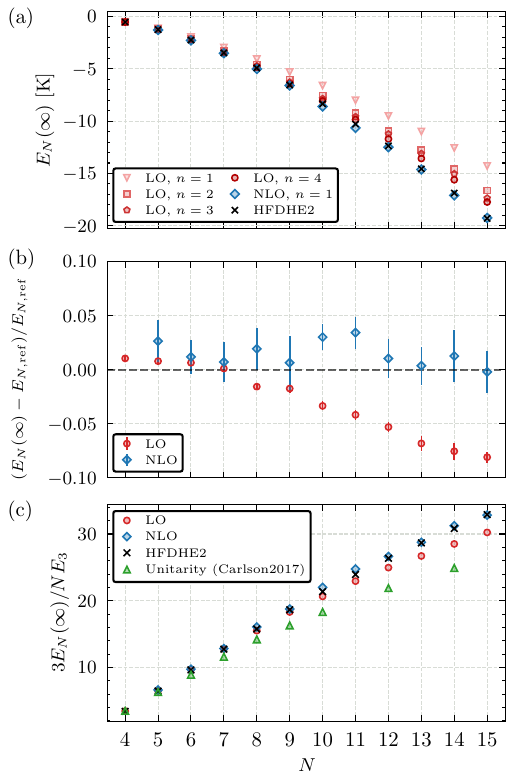}
    \caption{
    Dependence on the number $N$ of atoms in the cluster.
(a) Extrapolated energies $E_N(\infty)$ (in K) for LO with truncation orders $n=1$--4 and for NLO ($n=1$), compared with the HFDHE2 reference values.
(b) Relative deviation from reference $(E_N(\infty)-E_{N,\textrm{ref}})/E_{N,\textrm{ref}}$ for LO ($n=4$) and NLO.
(c) Trimer-scaled energy per particle $3E_N(\infty)/NE_3$  compared with HFDHE2 and  
unitarity \cite{Carlson:2017txq} results.
}
    \label{fig:Einf}
\end{figure}

Figure~\ref{fig:Einf}(a) shows that the LO extrapolation converges systematically with increasing truncation order $n$, while NLO improves the systematic trend and large-$N$ accuracy. The relative deviations of extrapolated energies, shown in Fig.~\ref{fig:Einf}(b), reveal that  
LO gives a slight overbinding 
for small clusters and  
underbinding that increases with $N$, as one might expect from the increase in $Q_N$.  
Surprisingly, NLO remains within a few percent of HFDHE2 and mildly overbound throughout.

In fact, the EFT expansion is converging much better than expected. The scale for ${\cal R}$ is set by the inverse van der Waals length, $r_{\rm vdW}=(mC_6/\hbar^2)^{1/4}/2$, where $C_6$ is the coefficient of the long-range $-C_6/r^6$ term of the reference potential and $a_0$ is the Bohr radius. For the HFDHE2 potential, $r_{\rm vdW}\simeq 5.08\, a_0$ \cite{Yates:2022mxk} and $Q_N r_{\rm vdW}/\hbar\simeq 1$ already at $N=10$. Yet, NLO corrections remain $\lesssim 10\%$ of the LO value for $N\leqslant 15$. The coefficients $\{c_1\}$ are of ${\cal O}(1)$ and decrease significantly at NLO. This is consistent with a breakdown scale set by $1/r_{\rm vdW}$ enhanced by a prefactor of a few. A better estimate requires higher orders. Although the error grows with $N$ in the region we considered, the effects of saturation at larger $N$ raise hopes for convergence in the bulk.

The large $^4$He scattering length 
implies $^4$He systems are near the two-body unitarity limit, where the trimer energy provides the dominant scale.
In Fig.~\ref{fig:Einf}(c) we plot 
the extrapolated values for the trimer-scaled energies per particle, $3E_N(\infty)/N E_3$, at LO and NLO, and compare them with the corresponding values obtained at unitarity in
Ref.~\cite{Carlson:2017txq}. The positive 
scattering length 
leads to slightly stronger binding than at unitarity, where the dimer lies at threshold; accordingly, the helium clusters lie above the unitary values. The difference is nevertheless modest, particularly for small $N$, indicating that helium clusters remain close to the unitary regime. This proximity suggests that an NLO treatment around the unitary limit~\cite{Wu:2023mhg,Wu:2026pjt}, similar to what is done for nucleons~\cite{Konig:2016utl,Konig:2019xxk}, could quantitatively bridge to realistic $^4$He energies.

\Sec{Conclusion} 
Our results establish a controlled EFT description of $^4$He clusters from few- to many-body regimes. They show that the dynamics of these systems is controlled by very few parameters. The proximity to unitarity suggests that extending the theory around the unitarity limit at NLO offers a promising route to quantitatively connect universal and realistic bosonic clusters~\cite{vanKolck:2017jon}.
With the LECs now determined at LO and NLO, the same interaction can be directly applied to larger systems, including bulk $^4$He~\cite{Kalos:1981zz,DeBruynOuboter:1987}, a direction we will pursue.

\begin{acknowledgments}
We thank Feng Wu and the participants of the ECT* workshop ``Universality in strongly-interacting systems: from QCD to atoms'' for fruitful discussions.
This material is based upon work supported by the U.S.\ Department of Energy,
Office of Science, Office of Nuclear Physics, under Award No.~DE-FG02-04ER41338.
We acknowledge INFN and ISCRA for awarding access to the LEONARDO supercomputer, owned by the EuroHPC Joint Undertaking, hosted by CINECA (Italy) through the allocations: IsCc7{\_}EFTANS, INF25{\_}monstre, INF26{\_}monstre.
\end{acknowledgments}


\clearpage

\pagebreak
\twocolumngrid
\begin{center}
\textbf{\large Supplemental Material: \TITLE}
\end{center}
\setcounter{equation}{0}
\setcounter{figure}{0}
\setcounter{table}{0}
\setcounter{page}{1}
\makeatletter
\renewcommand{\theequation}{S\arabic{equation}}
\renewcommand{\thefigure}{S\arabic{figure}}
\renewcommand{\thetable}{S\arabic{table}}
\renewcommand{\bibnumfmt}[1]{[S#1]}

In this Supplemental Material, we detail the matching of the EFT to the underlying potential, including the calibration of low-energy constants at each order. We further present the resulting predictions for cluster energies, their cutoff dependence, and the associated methodology.

\section{Two-body sector}

At low energies, two-body scattering is dominated by the $s$ wave. The $S$ matrix is characterized by the phase shift $\delta_0(k)$ as a function of the relative wave number $k$, which is amenable to
the effective range expansion~\cite{Bethe:1949yr},
\begin{equation}
\label{eq:delta}
k \cot \delta_0(k) = -\frac{1}{a}+\frac{r_0 k^2}{2} + \mathcal{O}(k^4).
\end{equation}
The scattering length $a$ and effective range $r_0$ are extracted from the low-energy scattering solutions of the two-body Schrödinger equation (for a pedagogical introduction, see Ref.~\cite{Macedo-Lima:2023fzp}). For the HFDHE2 potential we obtain $a = 234.91$ $a_0$ and $r_0=13.976$ $a_0$ (with $a_0$ the Bohr radius), consistent with Refs.~\cite{Janzen:1995,Yates:2022mxk}. 
The corresponding $S$ matrix, truncated at the effective-range term, exhibits a pole at an energy that differs by less than 0.1\% from the value obtained from
the solution of the Sch\"odinger equation for the dimer ground state with the same potential, $E_{\rm 2,ref}=-0.8348$ mK, which is also in agreement with Ref.~\cite{Janzen:1995}.

\subsection{Leading Order}

\begin{figure}[tb]
     \centering
     \includegraphics[scale = 1]{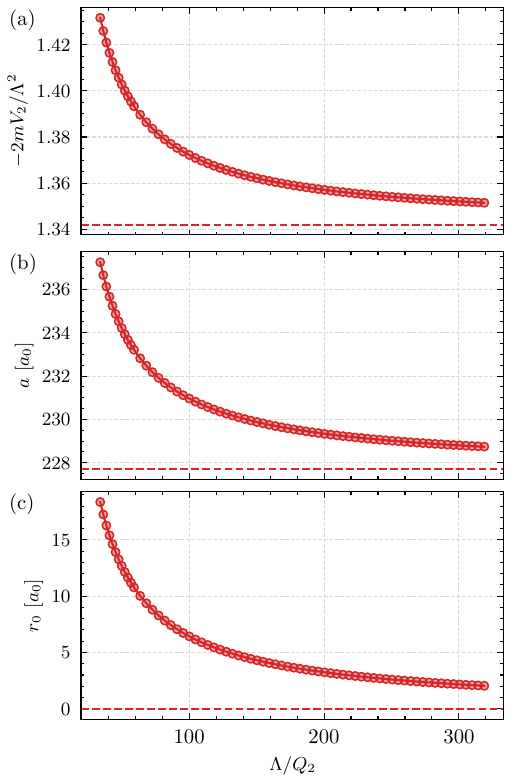}
     \caption{ 
     Dependence on the cutoff $\Lambda$ (in units of the characteristic dimer scale $Q_2$) of LO two-body quantities: (a) scaled LEC $-2mV^{\rm LO}_2/\Lambda^2$, fixed to reproduce the HFDHE2 dimer energy; (b) $s$-wave scattering length $a$ (in units of the Bohr radius $a_0$); and (c) effective range $r_0$ (in units of $a_0$). Curves are $n=2$ fits  
     to Eq.~(\ref{eq:ext_twobody}) for (a,b) and to Eq.~(\ref{eq:ext_r0}) for (c).
     The red dashed line indicates the infinite-cutoff extrapolation.}
     \label{fig:twobody_LO}
\end{figure}

At leading order (LO),
the low-energy constant (LEC) $V^{\rm LO}_2(\Lambda)$  is determined by reproducing the $^4$He dimer binding energy.
For each cutoff $\Lambda$, the two-body Schr\"odinger equation with the regulated LO interaction  $V_2^{\rm LO}(\Lambda) e^{-r^2 \Lambda^2/4\hbar^2}$ is solved numerically, and the LEC is adjusted until the ground-state energy matches $E_{\rm 2,\rm{ref}}$. The resulting cutoff dependence (``running'') of the scaled LEC $-2mV_2^{\rm LO}(\Lambda)/\Lambda^2$ is shown in Fig.~\ref{fig:twobody_LO}(a). It can be fitted with the analogue of Eq.~(\ref{eq:ext_powers}) of the main text for a generic two-body quantity $O$,
\begin{equation}
\label{eq:ext_twobody}
O(\Lambda)=O(\infty)\left[1+\sum_{i=1}^{n} d_i\left(\frac{Q_2}{\Lambda}\right)^i\right],
\end{equation}
where $O(\infty)$ denotes the cutoff-independent limit and $d_i$ are parameters. 
The fitted parameters are given in Table~\ref{tab:ext_LEC}. 
The potential is attractive and scales with the cutoff only as $\Lambda^2$, matching the singularity of the kinetic term and just overcoming its repulsive effect~\cite{Jackiw:1991je}.

\begin{table}[bt]
\centering
\caption{Cutoff-extrapolation parameters for the two-body LO and NLO LECs
obtained from 
fits with
Eq.~(\ref{eq:ext_twobody}).
}
\label{tab:ext_LEC}
\begin{tabular}{cccc}
\toprule
$O$ & $O(\infty)$ & $d_1$ & $d_2$ \\
\midrule
$-2mV^{\rm LO}_2/\Lambda^2$             & $1.342002(2)$ 			& $2.2406(1)$ 		& $1.456(3)$ \\
$2m Q_2 V_{2,0}^{\rm NLO}/\Lambda^3$    & $0.11020(4)$ 			    & $8.92(5)$ 		&  \\
$2m Q_2 V_{2,2}^{\rm NLO}/\Lambda^3$    & $-0.02756(1)$ 			& $10.14(5)$ 		& \\
\bottomrule
\end{tabular}
\end{table}

The corresponding LO scattering length and effective range are shown in Fig.~\ref{fig:twobody_LO}(b,c) as functions of the cutoff $\Lambda$. The scattering-length data can be 
fitted with Eq. \eqref{eq:ext_twobody}, but the 
effective range requires separate consideration, since it should vanish in the zero-range limit. Therefore, we analyze its cutoff dependence through
\begin{equation}
\label{eq:ext_r0}
O(\Lambda)=O(\infty)+d_1\frac{Q_2}{\Lambda}+d_2\left(\frac{Q_2}{\Lambda}\right)^2
+\ldots.
\end{equation}
Table~\ref{tab:ext_twobody_scattering} summarizes the cutoff extrapolations for $a$ and $r_0$ 
when Eqs. \eqref{eq:ext_twobody} and \eqref{eq:ext_r0} are truncated at $(Q_2/\Lambda)^2$.
The expansion coefficients are natural confirming that $Q_2$ is the scale that governs the two-body $^4$He system at distances larger than the van der Waals length. 
The extrapolated effective range is extremely small already with this truncation, indicating a dominant $\Lambda^{-1}$ dependence known from analytic calculations with separable regulators \cite{vanKolck:1998bw}.

\begin{table}[bt]
\centering
\caption{
Two-body scattering parameters at LO. The 
scattering length $a^{\rm LO}$ is fitted with Eq.~(\ref{eq:ext_twobody}), while the 
effective range $r_0^{\rm LO}$ is described by Eq.~(\ref{eq:ext_r0}).}
\label{tab:ext_twobody_scattering}
\begin{tabular}{cccc}
\toprule
$O$ & $O(\infty)$ & $d_1$ & $d_2$  \\
\midrule
$a^{\rm LO}/a_0$ 						& $227.72(2)$  	& $1.424(8)$ 		& $0.23(3)$ \\
$r_0^{\rm LO} Q_2/\hbar$ 						& 
$\; 9(1)\times 10^{-6}$\; & $2.868(3)$  	& $-3.904(7)$ \\
\bottomrule
\end{tabular}
\end{table}

The extrapolated scattering length given Table~\ref{tab:ext_twobody_scattering}
yields a zero-range dimer energy $-\hbar^2/m [a^{\rm LO}(\infty)]^2=-0.8346(1)$ mK. This energy is remarkably close to $E_{\rm 2,ref}$, consistently with the smallness of the constant term $d_0$. The renormalized Delta-function interaction at LO provides an accurate description of the two-body system. The residual $\Lambda^{-1}$ dependence 
indicates that a correction appears at NLO which provides a finite contribution to the effective-range parameters. 

\subsection{Next-to-Leading Order}

Next-to-leading order (NLO) corrections are small and amenable to 
distorted-wave Born approximation (DWBA)~\cite{Taylor:1972pty}. The NLO potential $V_{\rm 2B}^{\rm NLO}(r;\Lambda)$ endows the two-body scattering amplitude with energy dependence and a non-vanishing $r_0$.
Because range corrections are included only at NLO and are treated perturbatively, 
one avoids~\cite{vanKolck:1998bw} the inconsistencies ~\cite{Wigner:1955zz,Hammer:2010fw} that arise when attempting to incorporate effective-range effects nonperturbatively~\cite{Phillips:1996ae,Beane:1997pk}.
We determine the LEC $V^{\rm NLO}_{2,2}(\Lambda)$ so that $r_0$ is reproduced. Since the range correction affects the dimer energy, which we already fixed at LO, we introduce a second two-body LEC, $V^{\rm NLO}_{2,0}(\Lambda)$, to keep
the HFDHE2 dimer energy $E_{2,\rm ref}$ unchanged.

\begin{figure}[tb]
     \centering
     \includegraphics[scale = 1]{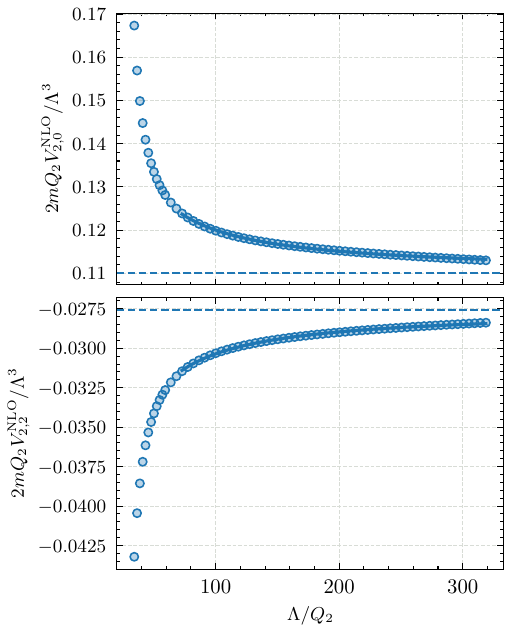}
     \caption{
     Dependence on the cutoff $\Lambda$ (in units of $Q_2$) of the NLO two-body LECs, $V^{\rm NLO}_{2,0}(\Lambda)$ (top panel) and $V^{\rm NLO}_{2,2}(\Lambda)$ (bottom panel). The LECs are fixed to reproduce the HFDHE2 dimer energy and effective range.
     The curves show $n=1$ fits to Eq.~(\ref{eq:ext_twobody}).
     The blue dashed lines indicate the infinite-cutoff extrapolations.
     }
     \label{fig:twobody_NLO}
\end{figure}

Figure~\ref{fig:twobody_NLO} shows the cutoff dependence of the two-body NLO LECs.
The extrapolation is described using Eq.~(\ref{eq:ext_twobody}) truncated at $n=1$, with the fitted parameters reported in Table~\ref{tab:ext_LEC}. The fact that both NLO LECs scale as 
$\Lambda^3$ for large cutoffs is consistent with previous works~\cite{vanKolck:1998bw}.
The negative values of $V^{\rm NLO}_{2,2}(\Lambda)$ are needed for $r_0>0$. The would-be increase in the dimer binding energy is compensated by a repulsive $V^{\rm NLO}_{2,0}(\Lambda)$.

\section{Three- and four-body sectors}

The three- and four-body $^4$He systems support bound states.
For the HFDHE2 potential, we obtain trimer and tetramer ground-state energies $E_{\rm 3,ref}=-0.1172(1)$ K and $E_{\rm 4,ref}=-0.534(1)$ K. These values are consistent with previous results, including Ref.~\cite{Pandharipande:1983xez}, which reports $-0.1173(3)$ K and $-0.533(2)$ K, and Ref.~\cite{Yates:2022mxk}, which finds $-0.1172(1)$ K and $-0.530(3)$ K.

Since an EFT incorporates the renormalization group, all interactions allowed by symmetries arise at some order in the EFT expansion. Renormalization provides a guide to determine at which order new interactions enter. At the orders new many-body interactions enter, new low-energy scales beyond $Q_2$ appear.

\subsection{
Thomas collapse}

\begin{figure}[tb]
     \centering
     \includegraphics[scale = 1]{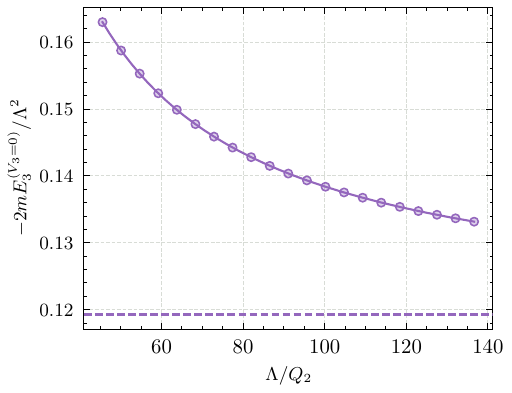}
     \caption{The Thomas collapse: scaled trimer ground-state energy  
     without a three-body force, $-2mE^{(V_{\rm 3B}=0)}_3(\Lambda)/\Lambda^2$, as a function of the cutoff $\Lambda$ (in units of $Q_2$). 
     The curve shows the $n=2$ fit to Eq.~(\ref{eq:ext_twobody}). 
     The black 
     dashed purple line indicates the extrapolated infinite-cutoff limit.}
     \label{fig:no3b}
\end{figure}

It is well known that the three-body system collapses under two-body zero-range interactions alone~\cite{Thomas:1935zz}. 
Under only the potential $V_{\rm 2B}^{\rm LO}(r;\Lambda)$,
the trimer ground-state energy
$E^{(V_{\rm 3B}=0)}_3(\Lambda)$ decreases
$\propto -\Lambda^2/2m$ as the cutoff increases. For the two-body force in Fig. \ref{fig:twobody_LO}(a), the cutoff dependence of the scaled energy $-2mE^{(V_{\rm 3B}=0)}_3(\Lambda)/\Lambda^2$ is shown in Fig.~\ref{fig:no3b}. It is well described by Eq.~(\ref{eq:ext_twobody}) truncated at $n=2$, with parameters listed in Table~\ref{tab:ext_thomas}. Note the large $d_{1,2}$ coefficients. The same ultraviolet collapse occurs for all larger clusters, underscoring the necessity of many-body interactions to achieve renormalization.

\begin{table}[bt]
\centering
\caption{
Fit parameters for the Thomas collapse of the
trimer energy with $V^{\rm LO}_{\rm 3B}=0$. The dimensionless quantity $-2m E^{(V_{\rm 3B}=0)}_3/\Lambda^2$ is fitted with Eq.~(\ref{eq:ext_twobody}) at $n=2$.}
\label{tab:ext_thomas}
\begin{tabular}{cccc}
\toprule
$O$ & $O(\infty)$ & $d_1$ & $d_2$  \\
\midrule
$-2m E^{(V_3=0)}_3/\Lambda^2$ 	& $1.193(1)$ & $15.5(2)$ & $54(1)$ \\
\bottomrule
\end{tabular}
\end{table}

\subsection{Leading Order}

The Thomas collapse is a manifestation of the fact that a three-body force is required at LO to renormalize the three-body system~\cite{Bedaque:1998kg,Bedaque:1998km}.
To the extent that the LO describes the two-$^4$He system, a three-body force is needed at the same order. In agreement with other arguments~\cite{Frederico:1999}, the LO three-body force $V_{\rm 3B}^{\rm LO}$ introduces a new scale, which determines the position of the Efimov tower of states \cite{Efimov:1970zz,Efimov:1971}. For separable regulators, the three-body LEC has a log-periodic dependence on the new scale
\cite{Bedaque:1998kg, Bedaque:1998km, Chen:2025rti, Chen:2025iqp}, which reflects an approximate discrete scale invariance.

\begin{figure}[tb]
     \centering
     \includegraphics[scale = 1]{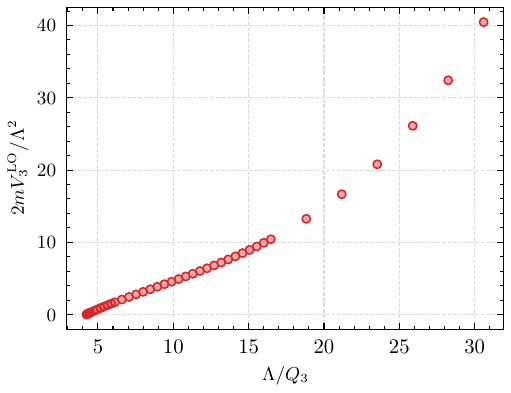}
     \caption{
     Dependence on the cutoff $\Lambda$ (in units of the characteristic trimer scale $Q_3$) of the LO three-body scaled LEC 
     $2m V^{\rm LO}_3(\Lambda)/\Lambda^2$, fixed to reproduce the HFDHE2 trimer energy.}
     \label{fig:three_body_LO}
\end{figure}

For each cutoff $\Lambda$, the three-body Schrödinger equation with the LO interaction is solved using diffusion Monte Carlo (DMC), and $V^{\rm LO}_3(\Lambda)$ is adjusted to match the $^4$He trimer ground-state energy obtained with the HFDHE2 potential, $E_{\rm 3,ref}$.
The resulting cutoff dependence of the LEC is shown in Fig.~\ref{fig:three_body_LO}. The three-body force is effectively repulsive to stabilize the Efimov spectrum, with a pronounced running that reflects the renormalization of the three-body counterterm for local Gaussian regulators \cite{Kirscher:2015yda}. 
In Fig.~\ref{fig:vr_levels_N4}(c) of the main text we show the convergence 
of the tetramer energy.

\subsection{Next-to-Leading Order}

The NLO two-body interactions change the three- and four-body energies. To keep the three-body energy fixed, an NLO correction to the three-body force is introduced. In contrast, a new, four-body force is required~\cite{Bazak:2018qnu} to eliminate the dramatic dependence of the four-body energy on the regulator.

\begin{figure}[tb]
     \centering
     \includegraphics[scale = 1]{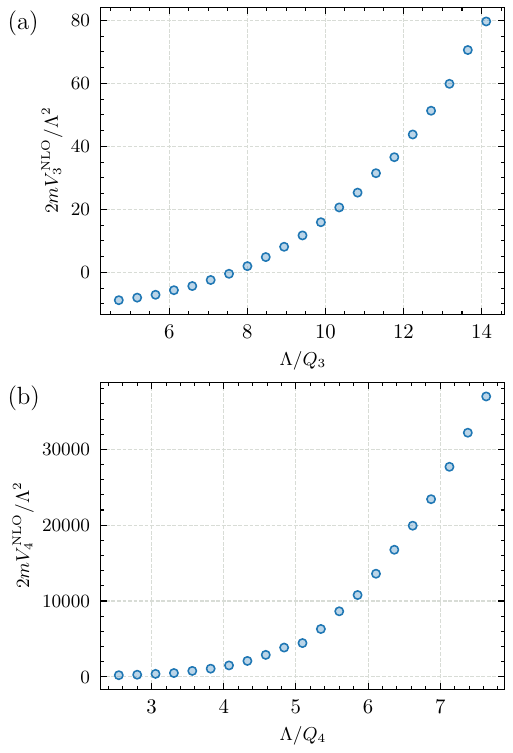}
     \caption{
     Scaled NLO three- and four-body LECs, determined by matching the HFDHE2 trimer and tetramer energies, respectively: (a) $2mV^{\rm NLO}_3(\Lambda)/\Lambda^2$ as a function of the cutoff $\Lambda$ (in units of $Q_3$); (b) $2mV^{\rm NLO}_4(\Lambda)/\Lambda^2$ as a function of $\Lambda$ (in units of the characteristic tetramer scale $Q_4$).}
     \label{fig:three_four_NLO}
\end{figure}

Thus,  
in addition to the two-body LECs determined within 
DWBA, three- and four-body LECs must also be fixed. The couplings $V^{\rm NLO}_3(\Lambda)$ and $V^{\rm NLO}_4(\Lambda)$ are constrained to reproduce the $^4$He trimer and tetramer ground-state energies from the HFDHE2 potential, $E_{\rm 3,ref}$ and $E_{\rm 4,ref}$.
The cutoff dependence of both LECs in shown in Fig.~\ref{fig:three_four_NLO}. For the smallest cutoffs considered, the NLO three-body force is attractive; as $\Lambda$ increases, it becomes repulsive. In contrast, the four-body force remains repulsive over the entire cutoff range.

\section{Clusters with 
$N\geqslant 5$}

The three- and four-body forces are known to be sufficient to guarantee the convergence of energies of up to six bosons at, respectively, LO \cite{Platter:2004he,Bazak:2016wxm} and NLO~\cite{Bazak:2018qnu}. An analytic argument based on the scaling of wave functions~\cite{Bazak:2018qnu} suggests that no more-body forces enter up to NLO, with a five-body force appearing at next-to-next-to-leading order for bosonic systems. Thus, the energies of clusters with $N\geqslant 5$ bosons can be predicted at LO and NLO. 

\begin{figure*}[htb]
     \centering
     \includegraphics[scale = 1.0]{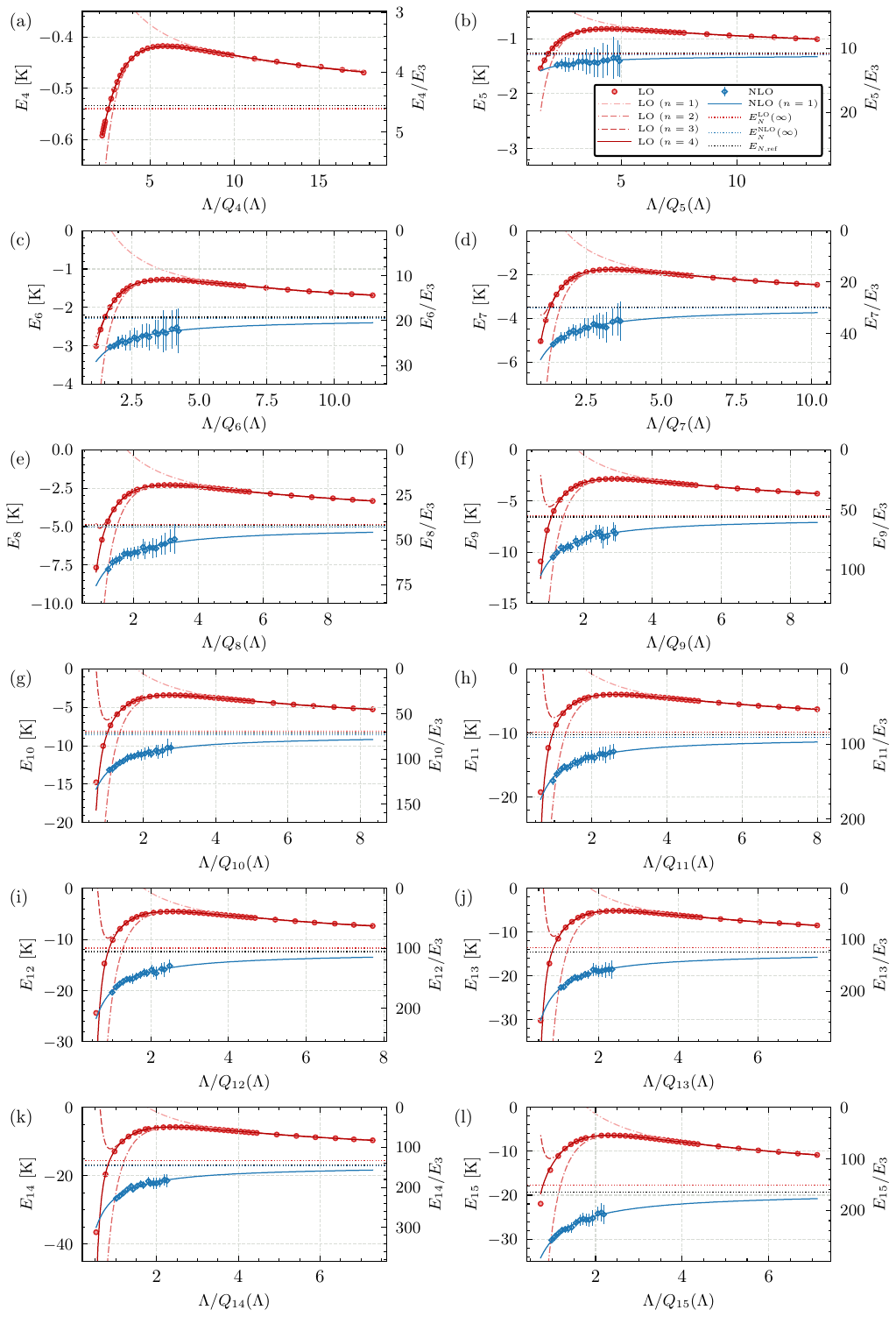}
     \caption{
     Dependence on the cutoff $\Lambda$ [through the dimensionless ratio $\Lambda/Q_N(\Lambda)$] of the ground-state energy $E_N$ (in K and in units of $E_3$) of clusters with $4\leqslant N \leqslant 15$ $^4$He atoms.
     Same plot elements as in Fig.~\ref{fig:N5_N10_N15} of the main text.
     }
     \label{fig:N4_to_N15}
\end{figure*}

In the main text, we presented representative results 
for $N=5$, 10, and 15 clusters [Fig.~\ref{fig:N5_N10_N15}(a-c)]. For completeness, Fig.~\ref{fig:N4_to_N15} shows the results for all cluster sizes in the range $4\leqslant N \leqslant 15$.

In the main text, the LO results were fitted to Eq.~(\ref{eq:ext_powers}) using a truncation order of $n=4$.
Here, we present the corresponding fits obtained by varying the truncation from $n=1$ to 4. Table~\ref{tab:extrapolated_LO} lists the coefficients $\{c_i^{\rm LO}\}$
together with the infinite-cutoff extrapolated energies $E^{\rm LO}_N(\infty)$. The latter are also expressed in units of the trimer energy and compared with the corresponding HFDHE2 values. Table~\ref{tab:extrapolated_NLO} presents the analogous results at NLO.

\begin{table*}[bt]
\centering
\caption{Leading-order cutoff extrapolations using Eq.~(\ref{eq:ext_powers}) for $4\leqslant N \leqslant 15$ and truncation orders $n=1$ to 4. The cutoff range employed in each fit is given by $[\Lambda/Q_N]_{\rm start}$ and $[\Lambda/Q_N]_{\rm end}$.}
\label{tab:extrapolated_LO}
\begin{tabular}{cccccccccc}
\toprule
$N$ & $[\Lambda/Q_N]_\mathrm{start}$ & $[\Lambda/Q_N]_\mathrm{end}$ & $c_1^{\rm LO}$ & $c_2^{\rm LO}$ & $c_3^{\rm LO}$ & $c_4^{\rm LO}$ & $E^{\rm LO}_N(\infty)$\! [K] & \, $E^{\rm LO}_N(\infty) / E_3$ \, & \, $E^{\rm LO}_N(\infty) / E_{N,\mathrm{ref}}$ \\
\midrule

4 & 7.990 & 17.652 & $-1.25(4)$ &  &  &  & $-0.501(2)$ & 4.27(2) & 0.937(5) \\
4 & 3.919 & 17.652 & $-2.45(2)$ & $7.10(8)$ &  &  & $-0.528(1)$ & 4.509(8) & 0.989(2) \\
4 & 2.176 & 17.652 & $-3.01(2)$ & $11.41(9)$ & $-9.6(1)$ &  & $-0.5426(9)$ & 4.630(7) & 1.016(2) \\
4 & 2.176 & 17.652 & $-2.90(6)$ & $10.5(4)$ & $-7(1)$ & $-3(1)$ & $-0.540(2)$ & 4.61(1) & 1.010(3) \\
\midrule

5 & 6.308 & 13.475 & $-1.58(4)$ &  &  &  & $-1.127(7)$ & 9.61(6) & 0.894(6) \\
5 & 3.383 & 13.475 & $-2.94(2)$ & $6.47(6)$ &  &  & $-1.225(3)$ & 10.46(2) & 0.972(2) \\
5 & 1.676 & 13.475 & $-3.39(2)$ & $9.52(7)$ & $-5.94(8)$ &  & $-1.259(2)$ & 10.74(2) & 0.998(2) \\
5 & 1.511 & 13.475 & $-3.55(4)$ & $10.6(3)$ & $-8.7(6)$ & $2.2(4)$ & $-1.271(4)$ & 10.84(4) & 1.008(4) \\
\midrule

6 & 5.641 & 11.411 & $-1.74(4)$ &  &  &  & $-1.96(1)$ & 16.8(1) & 0.870(7) \\
6 & 2.974 & 11.411 & $-3.08(2)$ & $5.75(5)$ &  &  & $-2.167(5)$ & 18.49(5) & 0.960(2) \\
6 & 1.522 & 11.411 & $-3.49(2)$ & $8.30(6)$ & $-4.52(7)$ &  & $-2.227(5)$ & 19.00(5) & 0.987(2) \\
6 & 1.183 & 11.411 & $-3.75(3)$ & $9.9(1)$ & $-8.2(2)$ & $2.6(1)$ & $-2.272(6)$ & 19.38(5) & 1.006(3) \\
\midrule

7 & 5.119 & 10.179 & $-1.80(4)$ &  &  &  & $-2.96(2)$ & 25.3(2) & 0.849(7) \\
7 & 2.944 & 10.179 & $-3.15(2)$ & $5.31(5)$ &  &  & $-3.317(8)$ & 28.30(7) & 0.952(2) \\
7 & 1.515 & 10.179 & $-3.55(2)$ & $7.67(6)$ & $-3.99(6)$ &  & $-3.415(8)$ & 29.14(7) & 0.980(2) \\
7 & 1.157 & 10.179 & $-3.81(3)$ & $9.3(1)$ & $-7.5(2)$ & $2.5(1)$ & $-3.49(1)$ & 29.78(9) & 1.001(3) \\
\midrule

8 & 4.767 & 9.370 & $-1.80(3)$ &  &  &  & $-4.07(3)$ & 34.8(3) & 0.827(6) \\
8 & 2.769 & 9.370 & $-3.14(1)$ & $4.88(4)$ &  &  & $-4.60(1)$ & 39.29(9) & 0.935(2) \\
8 & 1.402 & 9.370 & $-3.49(2)$ & $6.94(7)$ & $-3.38(7)$ &  & $-4.73(2)$ & 40.3(1) & 0.960(3) \\
8 & 1.033 & 9.370 & $-3.76(3)$ & $8.5(1)$ & $-6.6(2)$ & $2.1(1)$ & $-4.85(2)$ & 41.4(1) & 0.984(3) \\
\midrule

9 & 4.513 & 8.772 & $-1.82(3)$ &  &  &  & $-5.31(5)$ & 45.3(4) & 0.811(7) \\
9 & 2.643 & 8.772 & $-3.13(1)$ & $4.61(4)$ &  &  & $-6.05(2)$ & 51.7(1) & 0.924(2) \\
9 & 1.494 & 8.772 & $-3.56(2)$ & $6.89(6)$ & $-3.45(6)$ &  & $-6.29(2)$ & 53.6(1) & 0.959(2) \\
9 & 1.142 & 8.772 & $-3.82(3)$ & $8.4(1)$ & $-6.7(2)$ & $2.2(1)$ & $-6.44(3)$ & 54.9(2) & 0.983(4) \\
\midrule

10 & 4.320 & 8.344 & $-1.80(3)$ &  &  &  & $-6.61(5)$ & 56.4(5) & 0.792(6) \\
10 & 2.548 & 8.344 & $-3.09(1)$ & $4.32(3)$ &  &  & $-7.56(2)$ & 64.5(1) & 0.906(2) \\
10 & 1.438 & 8.344 & $-3.51(2)$ & $6.48(6)$ & $-3.13(6)$ &  & $-7.86(2)$ & 67.1(2) & 0.942(3) \\
10 & 1.086 & 8.344 & $-3.77(3)$ & $8.0(1)$ & $-6.1(2)$ & $2.0(1)$ & $-8.07(3)$ & 68.8(3) & 0.967(4) \\
\midrule

11 & 4.166 & 7.999 & $-1.80(3)$ &  &  &  & $-8.01(7)$ & 68.3(6) & 0.778(6) \\
11 & 2.472 & 7.999 & $-3.07(1)$ & $4.14(3)$ &  &  & $-9.21(2)$ & 78.6(2) & 0.896(2) \\
11 & 1.395 & 7.999 & $-3.49(2)$ & $6.23(6)$ & $-2.95(6)$ &  & $-9.59(3)$ & 81.8(3) & 0.932(3) \\
11 & 1.045 & 7.999 & $-3.76(3)$ & $7.7(1)$ & $-5.9(2)$ & $1.9(1)$ & $-9.86(5)$ & 84.1(4) & 0.958(4) \\
\midrule

12 & 4.150 & 7.716 & $-1.81(3)$ &  &  &  & $-9.50(8)$ & 81.1(7) & 0.769(6) \\
12 & 2.565 & 7.716 & $-3.07(1)$ & $4.03(3)$ &  &  & $-10.96(2)$ & 93.6(2) & 0.887(2) \\
12 & 1.202 & 7.716 & $-3.37(2)$ & $5.69(7)$ & $-2.49(6)$ &  & $-11.24(5)$ & 95.9(5) & 0.909(4) \\
12 & 1.014 & 7.716 & $-3.72(3)$ & $7.4(1)$ & $-5.5(2)$ & $1.7(1)$ & $-11.70(6)$ & 99.9(5) & 0.947(5) \\
\midrule

13 & 3.940 & 7.486 & $-1.79(3)$ &  &  &  & $-10.99(9)$ & 93.8(8) & 0.755(6) \\
13 & 2.514 & 7.486 & $-3.04(1)$ & $3.89(2)$ &  &  & $-12.77(2)$ & 109.0(2) & 0.877(2) \\
13 & 1.178 & 7.486 & $-3.34(2)$ & $5.50(7)$ & $-2.36(5)$ &  & $-13.10(6)$ & 111.8(5) & 0.900(4) \\
13 & 1.178 & 7.486 & $-3.64(6)$ & $6.9(3)$ & $-4.8(4)$ & $1.3(2)$ & $-13.6(1)$ & 115.8(9) & 0.932(7) \\
\midrule

14 & 3.852 & 7.288 & $-1.78(3)$ &  &  &  & $-12.6(1)$ & 107.3(9) & 0.745(6) \\
14 & 2.322 & 7.288 & $-3.01(1)$ & $3.72(3)$ &  &  & $-14.61(4)$ & 124.6(3) & 0.866(2) \\
14 & 1.158 & 7.288 & $-3.32(2)$ & $5.34(7)$ & $-2.26(6)$ &  & $-15.03(8)$ & 128.3(7) & 0.891(5) \\
14 & 1.158 & 7.288 & $-3.62(6)$ & $6.8(3)$ & $-4.6(5)$ & $1.3(3)$ & $-15.6(1)$ & 133(1) & 0.925(8) \\
\midrule

15 & 3.876 & 7.119 & $-1.79(3)$ &  &  &  & $-14.3(1)$ & 122(1) & 0.741(6) \\
15 & 2.430 & 7.119 & $-3.010(9)$ & $3.68(2)$ &  &  & $-16.62(3)$ & 141.8(3) & 0.862(2) \\
15 & 1.293 & 7.119 & $-3.41(1)$ & $5.58(4)$ & $-2.50(3)$ &  & $-17.36(4)$ & 148.2(4) & 0.900(2) \\
15 & 1.137 & 7.119 & $-3.60(4)$ & $6.5(2)$ & $-4.3(3)$ & $1.1(2)$ & $-17.7(1)$ & 151.2(9) & 0.919(5) \\

\bottomrule
\end{tabular}
\end{table*}

\begin{table*}[bt]
\centering
\caption{Same as Table~\ref{tab:extrapolated_LO}, but for NLO 
and 
$n=1$.}
\label{tab:extrapolated_NLO}
\begin{tabular}{ccccccc}
\toprule
$N$ & $[\Lambda/Q_N]_\mathrm{start}$ & $[\Lambda/Q_N]_\mathrm{end}$ & $c_1^{\rm NLO}$ & $E^{\rm NLO}_N(\infty)$\! [K] & \, $E^{\rm NLO}_N(\infty) / E_3$ \, & \, $E^{\rm NLO}_N(\infty) / E_{N,\mathrm{ref}}$ \\
\midrule

5 & 2.226 & 4.918 & $0.34(7)$ & $-1.29(2)$ & 11.0(2) & 1.03(2) \\

6 & 1.702 & 4.229 & $0.58(5)$ & $-2.28(4)$ & 19.5(3) & 1.01(2) \\

7 & 1.406 & 3.635 & $0.66(5)$ & $-3.51(6)$ & 29.9(5) & 1.01(2) \\

8 & 1.227 & 3.274 & $0.65(5)$ & $-5.0(1)$ & 42.8(8) & 1.02(2) \\

9 & 1.121 & 2.929 & $0.65(6)$ & $-6.6(2)$ & 56(1) & 1.01(2) \\

10 & 1.055 & 2.758 & $0.56(3)$ & $-8.6(1)$ & 73.3(9) & 1.03(1) \\

11 & 1.066 & 2.578 & $0.58(3)$ &  $-10.6(2)$ & 91(1) & 1.03(1) \\

12 & 1.001 & 2.483 & $0.61(4)$ & $-12.5(2)$ & 107(2) & 1.01(2) \\

13 & 1.058 & 2.340 & $0.58(4)$ & $-14.6(3)$ & 125(2) & 1.00(2) \\

14 & 1.011 & 2.254 & $0.55(5)$ & $-17.1(4)$ & 146(3) & 1.01(2) \\

15 & 1.062 & 2.189 & $0.57(4)$ & $-19.2(4)$ & 164(3) & 1.00(2) \\

\bottomrule
\end{tabular}
\end{table*}

\begin{figure}[tb]
     \centering
     \includegraphics[scale = 1]{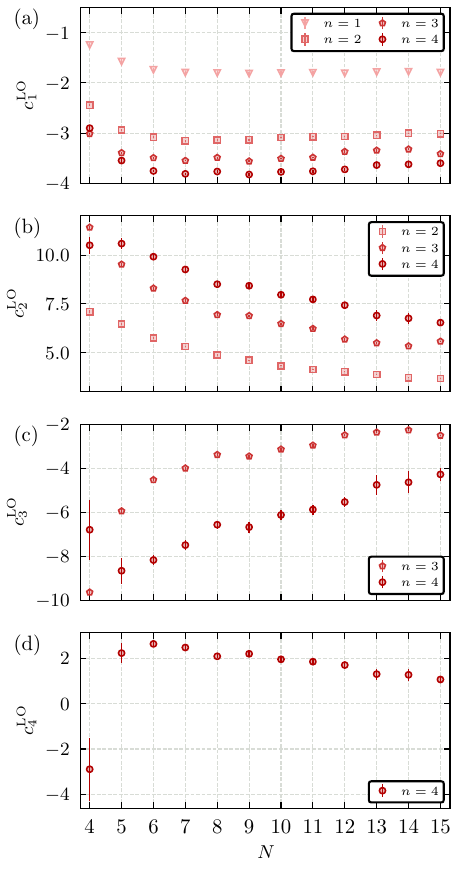}
     \caption{Coefficients $\{c_i^{\rm LO}\}$ extracted from fits of Eq.~(\ref{eq:ext_powers}) to the LO results for cluster sizes $4\leqslant N \leqslant 15$ and truncation orders $n=1$ to 4.}
     \label{fig:c1_to_c4}
\end{figure}

Figure~\ref{fig:c1_to_c4} summarizes the extracted coefficients $\{c_i^{\rm LO}\}$ for each truncation order and cluster size. The coefficient $c_1^{\rm LO}$, shown in Fig.~\ref{fig:c1_to_c4}(a), appears at all truncation orders considered and therefore provides a probe of the LO EFT expansion across cluster sizes. For fixed $N$, the variation of $c_1^{\rm LO}$ tends to decrease as the truncation order $n$ increases, indicating a convergent pattern. A similar behavior is observed for $c_2^{\rm LO}$ [Fig.~\ref{fig:c1_to_c4}(b)], although one fewer truncation order is available for comparison (by construction). 

The LO coefficients can also be analyzed as functions of the cluster size. Their variations are more pronounced for small systems, particularly for $N=4$, whereas for larger $N$ they approach a near-saturation behavior. This trend is reflected in Fig.~\ref{fig:N5_N10_N15}(d), where the $N=4$ curve exhibits a distinct cutoff dependence, while
curves lie 
closer together as $N$ increases.

\begin{figure}[tb]
     \centering
     \includegraphics[scale = 1]{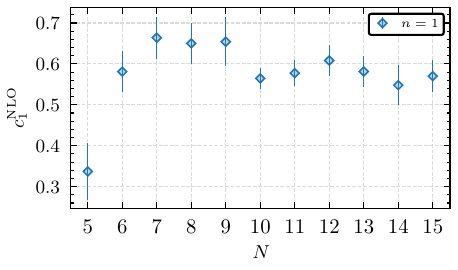}
     \caption{Coefficient $c_1^{\rm NLO}$ extracted from a fit of Eq.~(\ref{eq:ext_powers}) to the NLO results for cluster sizes $5\leqslant N \leqslant 15$ and truncation order $n=1$.}
     \label{fig:c1_NLO}
\end{figure}

Figure \ref{fig:c1_NLO} shows the coefficient $c_1^{\rm NLO}$ for the various cluster sizes we consider. One sees a distinct behavior at $N=5$ followed by gentler variation. Although the values of $c_1$ at LO change with $n$ by a factor of $\sim 2$, 
at NLO they are 
quite a bit smaller. This is expected since the $\Lambda^{-1}$ terms at one order in the EFT expansion are normally comparable to extrapolated values at the next order when the cutoff is on the order of the breakdown scale. If we compare $n=1$ truncations at the two orders, $c_1$ decreases by a factor $\sim 3$ across cluster sizes. In contrast, we would have expected this suppression factor to decrease with cluster size on account of the increasing value of $Q_N$.

\begin{thebibliography}{99}

\bibitem{Braaten:2004rn}
E.~Braaten and H.-W.~Hammer,
``Universality in few-body systems with large scattering length,''
Phys. Rept. \textbf{428} (2006) 259.

\bibitem{Naidon:2016dpf}
P.~Naidon and S.~Endo,
``Efimov Physics: a review,''
Rept. Prog. Phys. \textbf{80} (2017) 056001.

\bibitem{Chin:2010crf}
C.~Chin, R.~Grimm, P.~Julienne and E.~Tiesinga,
``Feshbach resonances in ultracold gases,''
Rev. Mod. Phys. \textbf{82} (2010) 1225.

\bibitem{Hammer:2019poc}
H.-W.~Hammer, S.~K{\"o}nig and U.~van Kolck,
``Nuclear effective field theory: status and perspectives,''
Rev. Mod. Phys. \textbf{92} (2020) 025004.

\bibitem{Kievsky:2021ghz}
A.~Kievsky, L.~Girlanda, M.~Gattobigio and M.~Viviani,
``Efimov Physics and Connections to Nuclear Physics,''
Ann. Rev. Nucl. Part. Sci. \textbf{71} (2021) 465.

\bibitem{Aziz:1979}
R.~A.~Aziz, V.~P.~S.~Nain, J.~S.~Carley, W.~L.~Taylor and G.~T.~McConville,
``An accurate intermolecular potential for helium,''
J. Chem. Phys. \textbf{70} (1979) 4330.

\bibitem{Aziz:1991}
R.~A.~Aziz and M.~J.~Slaman,
``An examination of ab initio results for the helium potential energy curve,''
J. Chem. Phys. \textbf{94} (1991) 8047.

\bibitem{Janzen:1995}
A.~R.~Janzen and R.~A.~Aziz,
``Modern He–He potentials: Another look at binding energy, effective range theory, retardation, and Efimov states,''
J. Chem. Phys. \textbf{103} (1995) 9626.

\bibitem{Janzen:1997}
A.~R.~Janzen and R.~A.~Aziz,
``An accurate potential energy curve for helium based on ab initio calculations,''
J. Chem. Phys. \textbf{107} (1997) 914.

\bibitem{Korona:1997}
T.~Korona, H.~L.~Williams, R.~Bukowski, B.~Jeziorski and K.~Szalewicz,
``Helium dimer potential from symmetry-adapted perturbation theory calculations using large gaussian geminal and orbital basis sets,''
J. Chem. Phys. \textbf{106} (1997) 5109.

\bibitem{Przybytek:2010zz}
M.~Przybytek, W.~Cencek, J.~Komasa, G.~Lach, B.~Jeziorski and K.~Szalewicz,
``Relativistic and Quantum Electrodynamics Effects in the Helium Pair Potential,''
Phys. Rev. Lett. \textbf{104} (2010) 183003
[erratum: Phys. Rev. Lett. \textbf{108} (2012) 129902].

\bibitem{Pandharipande:1983xez}
V.~R.~Pandharipande, J.~G.~Zabolitzky, S.~C.~Pieper, R.~B.~Wiringa and U.~Helmbrecht,
``Calculations of Ground-State Properties of Liquid $^4$He Droplets,''
Phys. Rev. Lett. \textbf{50} (1983) 1676.

\bibitem{Kievsky:2011ut}
A.~Kievsky, E.~Garrido, C.~Romero-Redondo and P.~Barletta,
``The helium trimer with soft-core potentials,''
Few Body Syst. \textbf{51} (2011) 259.

\bibitem{Hiyama:2011ge}
E.~Hiyama and M.~Kamimura,
``Variational calculation of 4He tetramer ground and excited states using a realistic pair potential,''
Phys. Rev. A \textbf{85} (2012) 022502.

\bibitem{Hiyama:2012cj}
E.~Hiyama and M.~Kamimura,
``Linear correlations between $^4$He trimer and tetramer energies calculated with various realistic $^4$He potentials,''
Phys. Rev. A \textbf{85} (2012) 062505.

\bibitem{Hiyama:2014kia}
E.~Hiyama and M.~Kamimura,
``Universality in Efimov-associated tetramers in $^4$He,''
Phys. Rev. A \textbf{90} (2014) 052514.

\bibitem{Kievsky:2014dua}
A.~Kievsky, N.~K.~Timofeyuk and M.~Gattobigio,
``$N$-boson spectrum from a Discrete Scale Invariance,''
Phys. Rev. A \textbf{90} (2014) 032504.

\bibitem{Kievsky:2020sni}
A.~Kievsky, A.~Polls, B.~Juli{\'a}-D{\'\i}az, N.~K.~Timofeyuk and M.~Gattobigio,
``Few bosons to many bosons inside the unitary window: A transition between universal and nonuniversal behavior,''
Phys. Rev. A \textbf{102} (2020) 063320.

\bibitem{Yates:2022mxk}
A.~J.~Yates and D.~Blume,
``Structural properties of $^4$He$_N$ (N=2{\textendash}10) clusters for different potential models at the physical point and at unitarity,''
Phys. Rev. A \textbf{105} (2022) 022824.

\bibitem{Recchia:2022jih}
P.~Recchia, A.~Kievsky, L.~Girlanda and M.~Gattobigio,
``Subleading contributions to N-boson systems inside the universal window,''
Phys. Rev. A \textbf{106} (2022) 022812.

\bibitem{Kalos:1981zz}
M.~H.~Kalos, M.~A.~Lee, P.~A.~Whitlock and G.~V.~Chester,
``Modern potentials and the properties of condensed He-4,''
Phys. Rev. B \textbf{24} (1981) 115.

\bibitem{Ceperley:1995zz}
D.~M.~Ceperley,
``Path integrals in the theory of condensed helium,''
Rev. Mod. Phys. \textbf{67} (1995) 279.

\bibitem{Moroni:2000}
S.~Moroni, F.~Pederiva, S.~Fantoni and M.~Boninsegni,
``Equation of state of solid $^3$He,''
Phys. Rev. Lett. \textbf{84} (2000) 2650.

\bibitem{Efimov:1970zz}
V.~Efimov,
``Energy levels arising from the resonant two-body forces in a three-body system,''
Phys. Lett. B \textbf{33} (1970) 563.

\bibitem{Efimov:1971}
V.~N.~Efimov,
``Weakly-bound states of 3 resonantly-interacting particles,''
Sov. J. Nucl. Phys. \textbf{12} (1971) 589.

\bibitem{Kunitski:2015qth}
M.~Kunitski, S.~Zeller, J.~Voigtsberger, A.~Kalinin, L.~P.~H.~Schmidt, M.~Sch{\"o}ffler, A.~Czasch, W.~Sch{\"o}llkopf, R.~E.~Grisenti and T.~Jahnke, \textit{et al.},
``Observation of the Efimov state of the helium trimer,''
Science \textbf{348} (2015) 551.

\bibitem{Bedaque:2002mn}
P.~F.~Bedaque and U.~van Kolck,
``Effective field theory for few-nucleon systems,''
Ann. Rev. Nucl. Part. Sci. \textbf{52} (2002) 339.

\bibitem{Bedaque:1998kg}
P.~F.~Bedaque, H.-W.~Hammer and U.~van Kolck,
``Renormalization of the three-body system with short-range interactions,''
Phys. Rev. Lett. \textbf{82} (1999) 463.

\bibitem{Bedaque:1998km}
P.~F.~Bedaque, H.-W.~Hammer and U.~van Kolck,
``The three-boson system with short-range interactions,''
Nucl. Phys. A \textbf{646} (1999) 444.

\bibitem{Bazak:2018qnu}
B.~Bazak, J.~Kirscher, S.~K{\"o}nig, M.~Pav{\'o}n Valderrama, N.~Barnea and U.~van Kolck,
``Four-Body Scale in Universal Few-Boson Systems,''
Phys. Rev. Lett. \textbf{122} (2019) 143001.

\bibitem{vanKolck:1998bw}
U.~van Kolck,
``Effective field theory of short-range forces,''
Nucl. Phys. A \textbf{645} (1999) 273.

\bibitem{Braaten:2002sr}
E.~Braaten, H.-W.~Hammer and M.~Kusunoki,
``Universal equation for Efimov states,''
Phys. Rev. A \textbf{67} (2003) 022505.

\bibitem{Braaten:2002jv}
E.~Braaten and H.-W.~Hammer,
``Universality in the three-body problem for He-4 atoms,''
Phys. Rev. A \textbf{67} (2003) 042706.

\bibitem{Platter:2004he}
L.~Platter, H.-W.~Hammer and U.-G.~Mei{\ss}ner,
``The four-boson system with short-range interactions,''
Phys. Rev. A \textbf{70} (2004) 052101.

\bibitem{Platter:2006ev}
L.~Platter and D.~R.~Phillips,
``The Three-Boson System at Next-To-Next-To-Leading Order,''
Few Body Syst. \textbf{40} (2006) 35.

\bibitem{Ji:2012nj}
C.~Ji and D.~R.~Phillips,
``Effective Field Theory Analysis of Three-Boson Systems at Next-To-Next-To-Leading Order,''
Few Body Syst. \textbf{54} (2013) 2317.

\bibitem{Bazak:2016wxm}
B.~Bazak, M.~Eliyahu and U.~van Kolck,
``Effective Field Theory for Few-Boson Systems,''
Phys. Rev. A \textbf{94} (2016) 052502.

\bibitem{Contessi:2023yoz}
L.~Contessi, M.~Sch{\"a}fer and U.~van Kolck,
``Improved action for contact effective field theory,''
Phys. Rev. A \textbf{109} (2024) 022814.

\bibitem{Wu:2023mhg}
F.~Wu, T.~Frederico, R.~Higa and U.~van Kolck,
``Four-boson first excited state near two-body unitarity,''
Phys. Rev. A \textbf{109} (2024) 043301.

\bibitem{Bazak:2025usn}
B.~Bazak,
``Small clusters of He atoms in finite-cutoff EFT,''
\href{https://arxiv.org/abs/2511.12538}{arXiv:2511.12538} [cond-mat.quant-gas].

\bibitem{Wu:2026pjt}
F.~Wu, X.~Lin, U.~van Kolck and S.~K{\"o}nig,
``Three- and four-boson systems expanded around the unitarity limit: Application to $^4$He,''
[arXiv:2606.00854 [cond-mat.quant-gas]]

\bibitem{Foulkes:2001zz}
W.~M.~C.~Foulkes, L.~Mitas, R.~J.~Needs and G.~Rajagopal,
``Quantum Monte Carlo simulations of solids,''
Rev. Mod. Phys. \textbf{73} (2001) 33.

\bibitem{Carlson:2017txq}
J.~Carlson, S.~Gandolfi, U.~van Kolck and S.~A.~Vitiello,
``Ground-state properties of unitary bosons: from clusters to matter,''
Phys. Rev. Lett. \textbf{119} (2017) 223002.

\bibitem{Madeira:2021qcz}
L.~Madeira, T.~Frederico, S.~Gandolfi, L.~Tomio and M.~T.~Yamashita,
``Quantum Monte Carlo studies of a trimer scaling function with microscopic two- and three-body interactions,''
Phys. Rev. A \textbf{104} (2021) 033301.

\bibitem{Grisenti:2000zz}
R.~E.~Grisenti, W.~Sch\"ollkopf, J.~P.~Toennies, G.~C.~Hegerfeldt, T.~Kohler and M.~Stoll,
``Determination of the Bond Length and Binding Energy of the Helium Dimer by Diffraction from a Transmission Grating,''
Phys. Rev. Lett. \textbf{85} (2000) 2284.

\bibitem{Voigtsberger:2014}
J.~Voigtsberger, S.~Zeller, J.~Becht, N.~Neumann, F.~Sturm, H.-K.~Kim, M.~Waitz,
F.~Trinter, M.~Kunitski, A.~Kalinin, J.~Wu, W.~Sch\"ollkopf, D.~Bressanini,
A.~Czasch, J.~B.~Williams, K.~Ullmann-Pfleger, L.~Ph.~H.~Schmidt,
M.~S.~Sch\"offler, R.~E.~Grisenti, T.~Jahnke and R.~D\"orner,
``Imaging the structure of the trimer systems
$^4$He$_3$ and $^3$He$^4$He$_2$,''
Nature Commun. \textbf{5} (2014) 5765.

\bibitem{Zeller:2016mwo}
S.~Zeller, M.~Kunitski, J.~Voigtsberger, A.~Kalinin, A.~Schottelius, C.~Schober, M.~Waitz, H.~Sann, A.~Hartung and T.~Bauer, \textit{et al.}
``Imaging the He$_2$ quantum halo state using a free electron laser,''
Proc. Natl. Acad. Sci. \textbf{113} (2016) 4651.

\bibitem{sup}
See Supplemental Material at [URL will be inserted by publisher] for details of the EFT calibration, extrapolation procedure, and additional cluster results.

\bibitem{Carleo:2016svm}
G.~Carleo and M.~Troyer,
``Solving the quantum many-body problem with artificial neural networks,''
Science \textbf{355} (2017) 602.

\bibitem{Lovato:2022tjh}
A.~Lovato, C.~Adams, G.~Carleo and N.~Rocco,
``Hidden-nucleons neural-network quantum states for the nuclear many-body problem,''
Phys. Rev. Res. \textbf{4} (2022) 043178.

\bibitem{Gnech:2021wfn}
A.~Gnech, C.~Adams, N.~Brawand, G.~Carleo, A.~Lovato and N.~Rocco,
``Nuclei with up to $A=6$ nucleons with artificial neural network wave functions,''
Few Body Syst. \textbf{63} (2022) 7.

\bibitem{Freitas:2023yed}
W.~Freitas and S.~A.~Vitiello,
``Synergy between deep neural networks and the variational Monte Carlo method for small $^4$He$_N$ clusters,''
Quantum \textbf{7} (2023) 1209.

\bibitem{Freitas:2024}
W.~Freitas, B.~Abreu and S.~A.~Vitiello,
``Modeling $^4$He$_N$ clusters with wave functions based on neural networks,''
J. Low Temp. Phys. \textbf{215} (2024) 357.

\bibitem{Konig:2016utl}
S.~K{\"o}nig, H.~W.~Grie{\ss}hammer, H.-W.~Hammer and U.~van Kolck,
``Nuclear Physics Around the Unitarity Limit,''
Phys. Rev. Lett. \textbf{118} (2017) 202501.

\bibitem{Konig:2019xxk}
S.~K{\"o}nig,
``Energies and radii of light nuclei around unitarity,''
Eur. Phys. J. A \textbf{56} (2020) 
113.

\bibitem{vanKolck:2017jon}
U.~van Kolck,
``Unitarity and Discrete Scale Invariance,''
Few-Body Syst. \textbf{58} (2017) 
112.

\bibitem{DeBruynOuboter:1987}
R.~De~Bruyn~Ouboter and C.~N.~Yang,
``The thermodynamic properties of liquid $^3$He--$^4$He mixtures between 0 and 20 atm in the limit of absolute zero temperature,''
Physica B+C \textbf{144} (1987) 127.

\bibitem{Bethe:1949yr}
H.~A.~Bethe,
``Theory of the Effective Range in Nuclear Scattering,''
Phys. Rev. \textbf{76} (1949) 38.

\bibitem{Macedo-Lima:2023fzp}
M.~Mac{\^e}do-Lima and L.~Madeira,
``Scattering length and effective range of microscopic two-body potentials,''
Rev. Bras. Ens. Fis. \textbf{45} (2023) e20230079.

\bibitem{Jackiw:1991je}
R.~Jackiw,
``Delta function potentials in two-dimensional and three-dimensional quantum mechanics,''
MIT-CTP-1937 (1991).

\bibitem{Taylor:1972pty}
J.~R.~Taylor,
``Scattering Theory: The Quantum Theory of Nonrelativistic Collisions,''
John Wiley {\&} Sons, Inc. (1972).

\bibitem{Wigner:1955zz}
E.~P.~Wigner,
``Lower Limit for the Energy Derivative of the Scattering Phase Shift,''
Phys. Rev. \textbf{98} (1955) 145.

\bibitem{Hammer:2010fw}
H.-W.~Hammer and D.~Lee,
``Causality and the effective range expansion,''
Annals Phys. \textbf{325} (2010) 2212.

\bibitem{Phillips:1996ae}
D.~R.~Phillips and T.~D.~Cohen,
``How short is too short? Constraining contact interactions in nucleon-nucleon scattering,''
Phys. Lett. B \textbf{390} (1997) 7.

\bibitem{Beane:1997pk}
S.~R.~Beane, T.~D.~Cohen and D.~R.~Phillips,
``The potential of effective field theory in NN scattering,''
Nucl. Phys. A \textbf{632} (1998) 445.

\bibitem{Thomas:1935zz}
L.~H.~Thomas,
``The Interaction Between a Neutron and a Proton and the Structure of H$^3$,''
Phys. Rev. \textbf{47} (1935) 903.

\bibitem{Frederico:1999}
T.~Frederico, L.~Tomio, A.~Delfino and A.~E.~A.~Amorim,
``Scaling limit of weakly bound triatomic states,''
Phys. Rev. A \textbf{60} (1999) R9.

\bibitem{Chen:2025rti}
L.~Chen and P.~Zhang,
``Exact renormalization relation and binding energies for three identical bosons,''
Phys. Rev. A \textbf{112} (2025) 
033319.

\bibitem{Chen:2025iqp}
L.~Chen, F.~Wu, X.~Lin, S.~K{\"o}nig, U.~van Kolck and P.~Zhang,
``Three-body limit cycle: Universal form for general regulators,''
Phys. Rev. A \textbf{113} (2026) 013314.

\bibitem{Kirscher:2015yda}
J.~Kirscher, N.~Barnea, D.~Gazit, F.~Pederiva and U.~van Kolck,
``Spectra and Scattering of Light Lattice Nuclei from Effective Field Theory,''
Phys. Rev. C \textbf{92} (2015) 
054002.


\end{thebibliography}
\end{document}